\documentclass[dvipsnames]{article}
\usepackage{spconf,amsmath,graphicx}
\usepackage{lipsum}
\usepackage{tikz}
\usetikzlibrary{arrows.meta, spy, calc, arrows}
\usepackage{adjustbox}
\usepackage[font=small,skip=0pt]{caption}
\usepackage{xcolor}
\usepackage{multirow}
\usepackage{floatrow}
\usepackage{array}
\usepackage{pgfplots}
\newcolumntype{P}[1]{>{\centering\arraybackslash}p{#1}}

\title{Multi-Structure Deep Segmentation with Shape Priors and Latent Adversarial Regularization}

\name{A. Boutillon$^{1,2}$ \qquad B. Borotikar$^{2,3,4}$ \qquad C. Pons$^{2,3,5}$ \qquad V. Burdin$^{1,2}$ \qquad P.-H. Conze$^{1,2}$ }

\address{$^{1}$ IMT Atlantique, Brest, France
         $^{2}$ LaTIM UMR 1101, Inserm, Brest, France \\
         $^{3}$ Centre Hospitalier R\'egional et Universitaire (CHRU) de Brest, Brest, France\\
         $^{4}$ Symbiosis Centre for Medical Image Analysis, Symbiosis International University, Pune, India\\
         $^{5}$ Fondation ILDYS, Brest, France}

\begin{document}

\maketitle

\begin{abstract}
\end{abstract}

Automatic segmentation of the musculoskeletal system in pediatric magnetic resonance (MR) images is a challenging but crucial task for morphological evaluation in clinical practice. We propose a deep learning-based regularized segmentation method for multi-structure bone delineation in MR images, designed to overcome the inherent scarcity and heterogeneity of pediatric data. Based on a newly devised shape code discriminator, our adversarial regularization scheme enforces the deep network to follow a learnt shape representation of the anatomy. The novel shape priors based adversarial regularization (SPAR) exploits latent shape codes arising from ground truth and predicted masks to guide the segmentation network towards more consistent and plausible predictions. Our contribution is compared to state-of-the-art regularization methods on two pediatric musculoskeletal imaging datasets from ankle and shoulder joints. 


\section{Introduction}
\label{sec:intro}

Semantic segmentation aims at partitioning an image into meaningful objects by localizing and extracting their boundaries. In the context of medical imaging, segmentation allows the generation of 3D models of anatomical structures which are then used to guide clinical decisions. For musculoskeletal analysis, it is essential to obtain the 3D shape of bones and muscles from MR images, as this information helps clinicians to diagnose pathologies, assess disease progression, monitor the effects of treatment or plan therapeutic interventions. While being the current standard for pediatric MR image delineation, manual segmentation poses limitations as it is laborious, tedious and subject to intra- and inter-observer variability. Thus, the use of 3D shape of bones and muscles as biomarkers is currently limited. Therefore, the development of robust and fully-automated segmentation techniques could benefit clinicians by reducing the analysis time and improving the reliability of morphological evaluation \cite{hirschmann_artificial_2019}.

Due to its promising performance, deep learning has recently become the predominant methodology for medical image analysis. The UNet architecture \cite{ronneberger_u-net_2015} is now considered as the baseline approach in the medical image segmentation community. Thus, UNet has already been applied for segmenting pediatric musculoskeletal structures such as shoulder muscles \cite{conze_healthy_2020} or ankle and shoulder bones \cite{boutillon_multi-structure_2020}. However, the development of automatic and reliable segmentation methods is hindered by the scarcity of medical imaging databases which is even more challenging for pediatric data. In the context of pathological pediatric examinations, the accuracy of multi-structure delineations is primordial in order to assess morphological changes which have a debilitating impact on child's growth \cite{balassy_role_2008}. Hence, developing segmentation techniques with enhanced accuracy and generalization abilities on small pediatric datasets is highly desirable.

Recent works focus on integrating regularization within neural networks through additional terms to the loss function to achieve higher generalizability. In the context of deep learning-based segmentation, regularization strategies can arise from different prior information: adversarial scheme \cite{singh_breast_2020}, shape priors \cite{oktay_anatomically_2018} or topological information \cite{clough_topological_2020}. Inspired by image-to-image translation, adversarial regularization aims at improving the plausibility of predicted segmentation masks. Alternatively, shape priors based regularization enforces the segmentation outputs to follow a learnt shape representation of the anatomy whereas topological regularization imposes output masks to comply with specified topological features. 

We proposed a new regularization scheme based on a non-linear shape representation learnt by an auto-encoder and a newly designed shape code discriminator trained in an adversarial fashion against the segmentation network. Our shape priors based adversarial contribution encouraged the segmentation network to follow global anatomical properties of the shape representation by guiding prediction masks closer to ground truth segmentation in latent shape space. We illustrated the effectiveness of our proposed approach for multi-bone segmentation on two scarce and heterogeneous pediatric musculoskeletal MR imaging datasets.

\begin{figure}[t]
\centering
\begin{adjustbox}{width=\textwidth}
\tikzstyle{dashed}=[dash pattern=on .9pt off .9pt]
\begin{tikzpicture}

\draw[line width=0.1mm, color=darkgray, rounded corners=1] (-.3, .3) -- (-.3,-1.35) -- (2.68,-1.35) -- (2.68,.3) -- cycle;
\draw[line width=0.1mm, color=darkgray, rounded corners=1] (-.3, -2) -- (-.3,-1.35) -- (2.68,-1.35) -- (2.68,-2) -- cycle;

\node[inner sep=0pt] (mri) at (0,0)
    {\includegraphics[width=.05\textwidth]{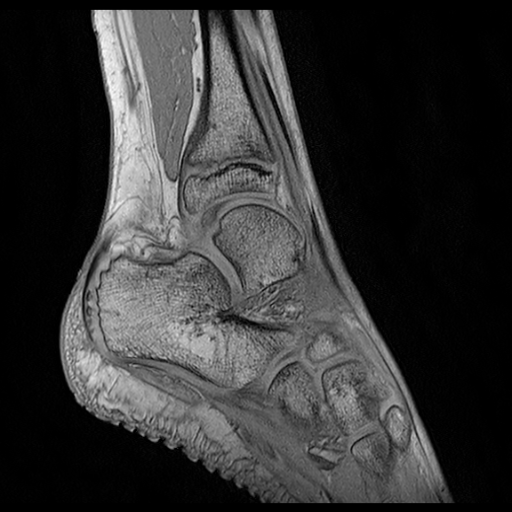}};
\node[inner sep=0pt] (pred) at (1.3,0)
    {\includegraphics[width=.05\textwidth]{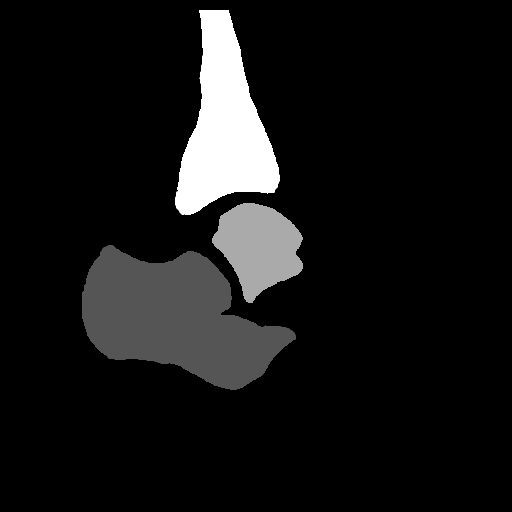}};
\node[inner sep=0pt] (gt) at (1.3,-.9)
    {\includegraphics[width=.05\textwidth]{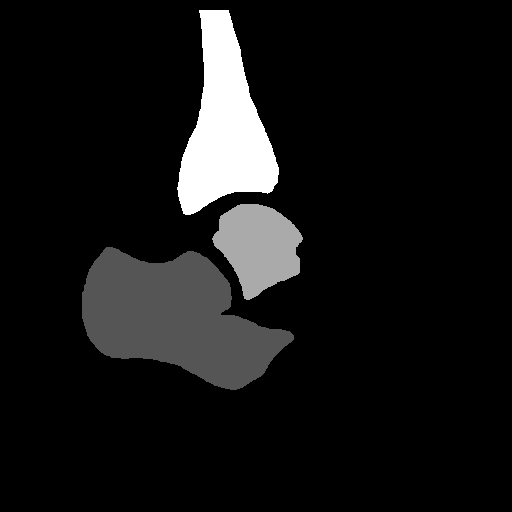}};

\draw[line width=0.01mm, fill=cyan!30] (.35,.25) -- (.65,.125) -- (.95,.25) -- (.95,-.25) -- (.65,-.125) -- (.35,-.25) -- cycle;
\node at (.65,0) {\scalebox{.23}{UNet}};

\draw[line width=0.01mm, fill=orange!40] (1.65,.25) -- (1.95,.125) -- (1.95,-.125) -- (1.65,-.25) -- cycle;
\draw[line width=0.01mm, fill=orange!40] (1.65,-.65) -- (1.95,-.775) -- (1.95,-1.025) -- (1.65,-1.15) -- cycle;
\node at (1.8,.04) {\scalebox{.21}{Shape}};
\node at (1.8,-.04) {\scalebox{.21}{encoder}};
\node at (1.8, -.86) {\scalebox{.21}{Shape}};
\node at (1.8,-.94) {\scalebox{.21}{encoder}};

\draw[line width=0.01mm, fill=Rhodamine!80] (2.15,-.2) -- (2.45,-.325) -- (2.45,-.575) -- (2.15,-.7) -- cycle;
\node at (2.3, -.37) {\scalebox{.21}{Shape}};
\node at (2.3, -.45) {\scalebox{.21}{Code}};
\node at (2.3, -.53) {\scalebox{.21}{Discr.}};

\draw[line width=0.01mm] (1.425,-.4) rectangle (1.175,-.5)  node[pos=.5] {\scalebox{.24}{{$\ell_{CE}$}}};
\draw[line width=0.01mm] (2.45,-.325) rectangle (2.55,-.575) node[pos=.5, rotate=90] {\scalebox{.24}{{$\ell_{SPAR}$}}};

\draw[line width=0.01mm, -{Latex[length=1.5pt, width=1.5pt]}] (mri.east) -- (.35,0);
\draw[line width=0.01mm, -{Latex[length=1.5pt, width=1.5pt]}] (.95,0) -- (pred.west);
\draw[line width=0.01mm, -{Latex[length=1.5pt, width=1.5pt]}] (pred.east) -- (1.65,0);
\draw[line width=0.01mm] (pred.south) -- (1.3,-.24);
\draw[line width=0.01mm, -{Latex[length=1.5pt, width=1.5pt]}] (1.3,-.32) -- (1.3,-.4);
\draw[line width=0.01mm, -{Latex[length=1.5pt, width=1.5pt]}] (gt.north) -- (1.3,-.5);
\draw[line width=0.01mm, -{Latex[length=1.5pt, width=1.5pt]}] (gt.east) -- (1.65,-.9);

\draw[line width=0.01mm, -{Latex[length=1.5pt, width=1.5pt]}] (1.95,0) -- (2.05,0) --(2.05,-.35) -- (2.15, -.35);
\draw[line width=0.01mm, -{Latex[length=1.5pt, width=1.5pt]}] (1.95,-.9) -- (2.05, -.9) -- (2.05,-.55) -- (2.15, -.55);

\draw[line width=0.01mm, dashed, -{Latex[length=1.5pt, width=1.5pt]}] (2.55, -0.45) -- (2.6, -0.45) -- (2.6, -1.25) -- (2.275,-1.25) -- (.65,-1.25) -- (.65,-.525);

\draw[line width=0.01mm, dashed, -{Latex[length=1.5pt, width=1.5pt]}] (1.175, -0.45) -- (.725, -0.45);

\draw[line width=0.01mm,dashed, -{Latex[length=1.5pt, width=1.5pt]}] (.65, -.375) -- (.65, -.125);

\draw[line width=0.01mm] (.65,-.45) circle (.075);
\draw[line width=0.01mm] (.65,-.4) -- (.65,-.5);
\draw[line width=0.01mm] (.6,-.45) -- (.7,-.45);

\node at (0,-.28) {\scalebox{.25}{MRI}};
\node at (1.3,-.28) {\scalebox{.25}{Prediction}};
\node at (1.3,-1.18) {\scalebox{.25}{Ground truth}};

\draw[line width=0.01mm, -{Latex[length=1.5pt, width=1.5pt]}] (-.17,-.9) -- (.03, -.9);
\draw[line width=0.01mm, dashed, -{Latex[length=1.5pt, width=1.5pt]}] (-.17,-1.15) -- (.03, -1.15);

\node[anchor=west] at (-.3, -.43) {\scalebox{.25}{\textbf{Training with}}};
\node[anchor=west] at (-.3, -.53) {\scalebox{.25}{\textbf{shape priors based}}};
\node[anchor=west] at (-.3, -.62) {\scalebox{.25}{\textbf{adversarial}}};
\node[anchor=west] at (-.3, -.73) {\scalebox{.25}{\textbf{regularization}}};

\node[anchor=west] at (-.07, -.85) {\scalebox{.25}{Forward}};
\node[anchor=west] at (-.07, -.95) {\scalebox{.25}{propagation}};
\node[anchor=west] at (-.07, -1.1) {\scalebox{.25}{Backward}};
\node[anchor=west] at (-.07, -1.2) {\scalebox{.25}{propagation}};

\draw[line width=0.01mm, fill=black] (.65,-1.45) rectangle (.7,-1.85);
\draw[line width=0.01mm, fill=black] (.8,-1.5) rectangle (.84,-1.8);
\draw[line width=0.01mm, fill=black] (.94,-1.55) rectangle (.97,-1.75);
\draw[line width=0.01mm, fill=black] (1.07,-1.6) rectangle (1.09,-1.7);

\draw[line width=0.01mm, fill=Lavender] (.72,-1.69) -- (.72,-1.61) -- (.78,-1.65) -- cycle;
\draw[line width=0.01mm, fill=Lavender] (.86,-1.69) -- (.86,-1.61) -- (.92,-1.65) -- cycle;
\draw[line width=0.01mm, fill=Lavender] (.99,-1.69) -- (.99,-1.61) -- (1.05,-1.65) -- cycle;
\draw[line width=0.01mm, fill=magenta] (1.11,-1.69) -- (1.11,-1.61) -- (1.17,-1.65) -- cycle;

\node at (0.675,-1.91) {\scalebox{.25}{512}};
\node at (0.82,-1.86) {\scalebox{.25}{256}};
\node at (0.955,-1.81) {\scalebox{.25}{128}};
\node at (1.08,-1.76) {\scalebox{.25}{64}};

\node[anchor=west] at (1.09,-1.58) {\scalebox{.25}{Fake 0}};
\node[anchor=west] at (1.09,-1.72) {\scalebox{.25}{Real 1}};

\node[anchor=west] at (-.3, -1.575) {\scalebox{.25}{\textbf{Shape code}}};
\node[anchor=west] at (-.3, -1.675) {\scalebox{.25}{\textbf{discriminator}}};
\node[anchor=west] at (-.3, -1.775) {\scalebox{.25}{\textbf{architecture}}};

\draw[line width=0.01mm, fill=Lavender] (1.65,-1.615) -- (1.65,-1.535) -- (1.71,-1.575) -- cycle;
\draw[line width=0.01mm, fill=magenta] (1.65,-1.815) -- (1.65,-1.735) -- (1.71,-1.775) -- cycle;

\node[anchor=west] at (1.62,-1.5135) {\scalebox{.25}{conv 1$\times$1, BN, ReLU}};
\node[anchor=west] at (1.62,-1.615) {\scalebox{.25}{max-pooling 2$\times$2}};
\node[anchor=west] at (1.62,-1.775) {\scalebox{.25}{conv 1$\times$1, sigmoid}};

\end{tikzpicture}
\end{adjustbox}
  \caption{SPAR method: the training procedure is based on UNet (top), a shape representation learnt by an auto-encoder and a shape code discriminator (bottom). The segmentation network exploits cross-entropy loss $\ell_{CE}$ and shape priors based adversarial regularization $\ell_{SPAR}$.}
  \label{fig:fig_framework}
\end{figure}
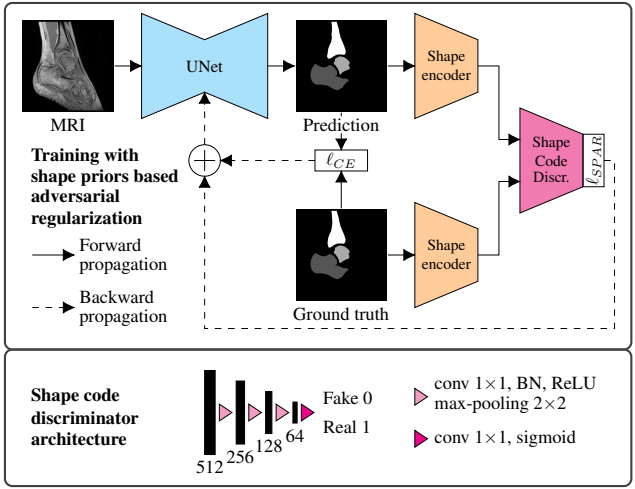

\section{Method}
\label{sec:method}

\subsection{Baseline segmentation model}
\label{sec:baseline}

Let $x$ be a greyscale image and $y$ the corresponding image of class labels. The mapping between intensities and class label images is approximated by a neural network $S : x \mapsto S(x)$ whose weights are learnt by optimizing the loss function $\ell_{S}$. Let $\hat{y} = S(x)$ be the estimate of $y$ having observed $x$ and $C$ the number of segmentation classes. We employed a categorical cross-entropy loss $\ell_{CE}(\hat{y}, y) = - \sum_{c=1}^{C} y_c \log(\hat{y}_c)$ to train the segmentation model through back-propagation. 

The architecture of $S$ was based on UNet \cite{ronneberger_u-net_2015}: a convolutional encoder-decoder with skip connections. Network layers were composed of a set of convolutional filters, batch normalization (BN), ReLU activation function, max-pooling and deconvolutional filters. A final softmax activation layer provided the predicted segmentation mask.

\subsection{Learning shape priors}
\label{sec:learning_shape}

Due to the constrained nature of anatomical structures in medical images, one can learn a compact representation of the anatomy arising from ground truth segmentation masks using an auto-encoder \cite{oktay_anatomically_2018}. An auto-encoder is composed of two sub-networks: an encoder $F : y \mapsto F(y)$ which produces a latent representation of the input, and a decoder $G : F(y) \mapsto G(F(y))$ which reconstructs the original data. The training procedure encourages the auto-encoder to reconstruct its input by optimizing the categorical cross-entropy between input and output $\ell_{CE}(G(F(y)), y)$. Consequently, the auto-encoder learned a non-linear compact representation of the shape, as the latent space was of lower dimensionality than the data space. Following the auto-encoder optimization, we employed the encoder $F$ with fixed weights to obtain the latent representation of both ground truth $y$ and predicted $\hat{y}$ masks. The shape encoder $F$ was composed of a set of convolutional layers, BN, ReLU activation and max-pooling layers. In the following, we note $z = F(y)$ and $\hat{z} = F(\hat{y})$ the latent shape codes arising from ground truth and estimated masks. 

\subsection{Shape priors based adversarial regularization (SPAR)}
\label{sec:shape_priors_adv_adversarial}

In the context of medical image segmentation, a discriminator typically assesses the plausibility of a given segmentation mask. Thus, incorporating an adversarial regularization in the segmentation network optimization scheme encourages the model to fool the discriminator and produce more plausible contours \cite{singh_breast_2020}. We adapted this approach to the latent shape representation arising from segmentation masks instead of segmentation masks themselves as usually done \cite{boutillon_multi-structure_2020}. Specifically, we proposed to adopt an adversarial training procedure similar to \cite{makhzani_adversarial_2016}, which encourages the predicted latent codes to be close to ground truth ones in the low dimensional shape space. Hence, the network learns to produce segmentation masks that match the learnt non-linear shape representation. Our approach is akin to \cite{oktay_anatomically_2018} which minimizes the Euclidean distance between predicted and ground truth latent codes.  

We designed a novel shape code discriminator $D : z \mapsto \{0,1\}$ which assess if an input latent shape code $z$ corresponds to a synthetic or real delineation mask. The architecture of such discriminator consisted of a succession of convolutional filters with 1$\times$1 kernel to reduce the feature dimensionality, followed by BN, max-pooling and ReLU (Fig.\ref{fig:fig_framework}). The number of features as well as the spatial dimensions were halved by two after each layer and a final sigmoid activation layer computed the likelihood of $z$ being fake ($0$) or real ($1$).

The training scheme optimized the shape code discriminator and segmentation network alternatively. The shape code discriminator was trained to differentiate latent codes using binary cross-entropy $\ell_{D} = - \log(1 - D(\hat{z})) - \log(D(z))$. At the segmentation network optimization step, the shape code discriminator computed the shape priors based adversarial regularization term $\ell_{SPAR}(\hat{y}) = - \log(D(F(\hat{y})))$ which represented the probability that the network considered the generated shape codes to be ground truth latent codes. Thus, the segmentation training strategy was modified as follows: $\ell_{S} = \ell_{CE}(\hat{y}, y) + \lambda \ell_{SPAR}(\hat{y})$ with $\lambda$ a weighting hyper-parameter. The optimization of $\ell_{SPAR}$ enforced $S$ to fool the discriminator and generate delineations whose latent representation was close to the ground truth shape representation.

\begin {figure*}[t]
\centering
\begin{adjustbox}{width=\textwidth}
\begin{tikzpicture}
\begin{scope}[spy using outlines=
      {circle, magnification=2.5, size=.4cm, connect spies, rounded corners}]

\node[inner sep=0pt] at (0,0)
    {\includegraphics[width=.082\textwidth]{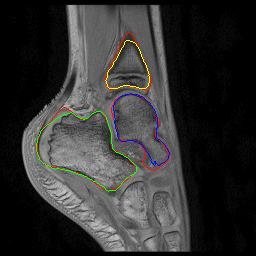}};
\node[inner sep=0pt] at (0,-1.5)
    {\includegraphics[width=.082\textwidth]{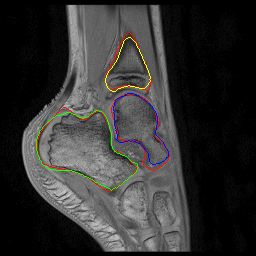}};
\node[inner sep=0pt] at (4.7,0)
    {\includegraphics[width=.082\textwidth]{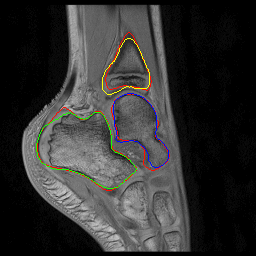}};
\node[inner sep=0pt] at (4.7,-1.5)
    {\includegraphics[width=.082\textwidth]{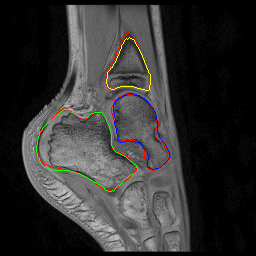}};

\node[inner sep=0pt] at (1.5,0)
    {\includegraphics[width=.082\textwidth]{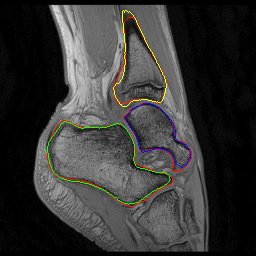}};
\node[inner sep=0pt] at (1.5,-1.5)
    {\includegraphics[width=.082\textwidth]{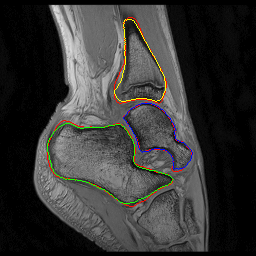}};
\node[inner sep=0pt] at (6.2,0)
    {\includegraphics[width=.082\textwidth]{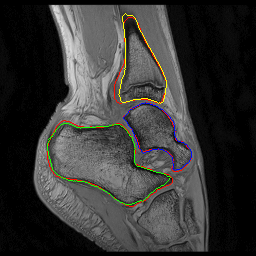}};
\node[inner sep=0pt] at (6.2,-1.5)
    {\includegraphics[width=.082\textwidth]{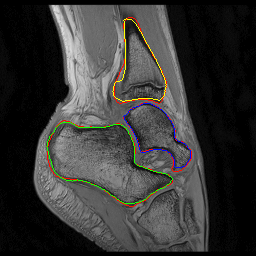}};
 
\node[inner sep=0pt] at (3,0)
    {\includegraphics[width=.082\textwidth]{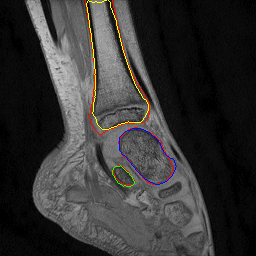}};
\node[inner sep=0pt] at (3,-1.5)
    {\includegraphics[width=.082\textwidth]{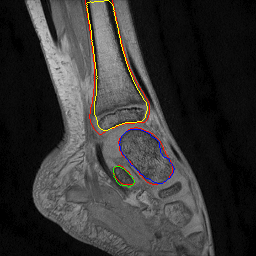}};
\node[inner sep=0pt] at (7.7,0)
    {\includegraphics[width=.082\textwidth]{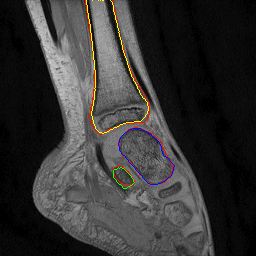}};
\node[inner sep=0pt] at (7.7,-1.5)
    {\includegraphics[width=.082\textwidth]{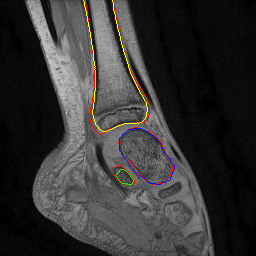}};
    
\spy [Dandelion] on (-0.36,.08) in node [left] at (-.24,.44);
\spy [Dandelion] on (-0.36,-1.42) in node [left] at (-.24,-1.06);
\spy [Dandelion] on (4.34,.08) in node [left] at (4.46,.44);
\spy [Dandelion] on (4.34,-1.42) in node [left] at (4.46,.-1.06);

\spy [Dandelion] on (0.1,.26) in node [left] at (.64,.44);
\spy [Dandelion] on (0.1,-1.24) in node [left] at (.64,-1.06);
\spy [Dandelion] on (4.8,.26) in node [left] at (5.34,.44);
\spy [Dandelion] on (4.8,-1.24) in node [left] at (5.34,-1.06);

\spy [Dandelion] on (0.14,-0.2) in node [left] at (.64,-.44);
\spy [Dandelion] on (0.14,-1.7) in node [left] at (.64,-1.94);
\spy [Dandelion] on (4.84,-.2) in node [left] at (5.34,-.44);
\spy [Dandelion] on (4.84,-1.7) in node [left] at (5.34,-1.94);

\spy [Dandelion] on (1.46,.15) in node [left] at (1.26,.44);
\spy [Dandelion] on (1.46,-1.35) in node [left] at (1.26,-1.06);
\spy [Dandelion] on (6.16,.15) in node [left] at (5.96,.44);
\spy [Dandelion] on (6.16,-1.35) in node [left] at (5.96,-1.06);

\spy [Dandelion] on (1.77,-.20) in node [left] at (2.14,.44);
\spy [Dandelion] on (1.77,-1.7) in node [left] at (2.14,-1.06);
\spy [Dandelion] on (6.47,-.2) in node [left] at (6.84,.44);
\spy [Dandelion] on (6.47,-1.7) in node [left] at (6.84,-1.06);

\spy [Dandelion] on (2.84,.01) in node [left] at (2.76,.44);
\spy [Dandelion] on (2.84,-1.49) in node [left] at (2.76,-1.06);
\spy [Dandelion] on (7.54,.01) in node [left] at (7.46,.44);
\spy [Dandelion] on (7.54,-1.49) in node [left] at (7.46,-1.06);

\spy [Dandelion] on (3.22,-.21) in node [left] at (3.64,.44);
\spy [Dandelion] on (3.22,-1.71) in node [left] at (3.64,-1.06);
\spy [Dandelion] on (7.92,-.21) in node [left] at (8.34,.44);
\spy [Dandelion] on (7.92,-1.71) in node [left] at (8.34,-1.06);

\spy [Dandelion] on (2.97,-.27) in node [left] at (2.76,-.44);
\spy [Dandelion] on (2.97,-1.77) in node [left] at (2.76,-1.94);
\spy [Dandelion] on (7.67,-.27) in node [left] at (7.46,-.44);
\spy [Dandelion] on (7.67,-1.77) in node [left] at (7.46,-1.94);

\end{scope}
    
\node[rotate=90] at (-.85, 0) {\scalebox{.5}{Base. UNet}};
\node[rotate=90] at (-.85, -1.5) {\scalebox{.5}{Adv. Reg.}};
\node[rotate=90] at (3.85, 0) {\scalebox{.5}{Sh. Reg.}};
\node[rotate=90] at (3.85, -1.5) {\scalebox{.5}{SPAR}};

\end{tikzpicture}
\end{adjustbox}
\caption{Visual comparison of UNet regularization methods: baseline UNet \cite{ronneberger_u-net_2015}, adversarial regularization \cite{singh_breast_2020}, shape priors based regularization \cite{oktay_anatomically_2018} and the proposed shape priors based adversarial regularization on ankle dataset. Ground truth delineations are in red (\textcolor{red}{\---}). Predicted bones, calcaneus, talus and tibia respectively appear in green (\textcolor{green}{\---}), blue (\textcolor{blue}{\---}) and yellow (\textcolor{yellow}{\---}).}
\label{fig:segmentation}
\end{figure*}
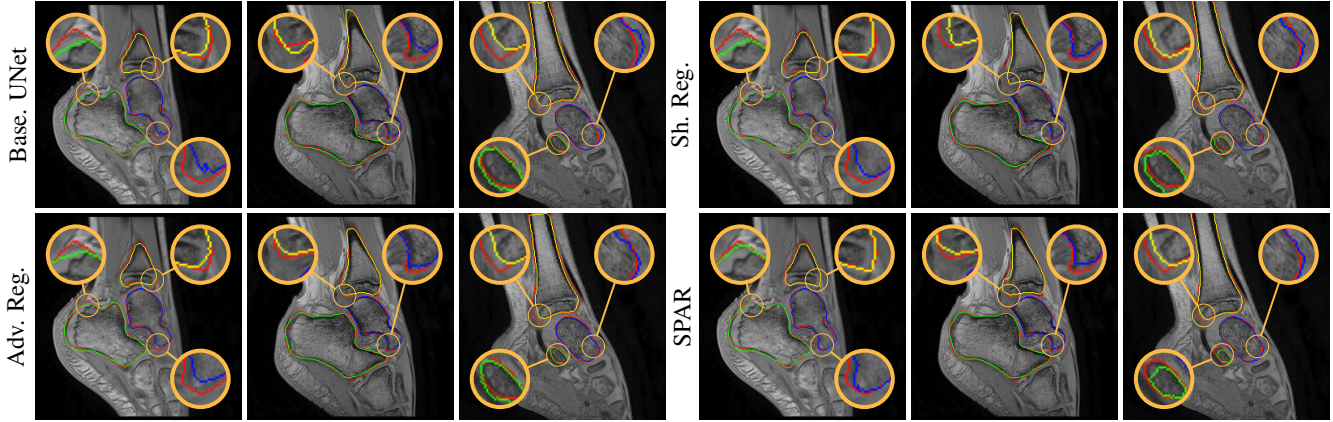

\section{Experiments}
\label{sec:experiments}

\subsection{Imaging datasets}
\label{sec:datasets}

Experiments were conducted on two pediatric datasets previously acquired using a 3T Philips scanner \cite{boutillon_multi-structure_2020}. The two MR images datasets were independently acquired on two musculoskeletal joints (ankle, shoulder) from a cohort of 17 and 15 pediatric patients. An expert (12 years of experience) annotated images to get ground truth contours of calcaneus, talus and tibia for ankle, as well as scapula and humerus for shoulder. All axial slices were downsampled to 256$\times$256 pixels.

\subsection{Implementation details}
\label{sec:implementation}

We compared the proposed shape priors based adversarial regularization method (SPAR in Fig.\ref{fig:fig_framework}) with baseline UNet (Base. UNet) \cite{ronneberger_u-net_2015}, adversarial regularization (Adv. Reg.) \cite{singh_breast_2020} and shape priors based regularization (Sh. Reg.) \cite{oktay_anatomically_2018}. For all methods, the backbone UNet architecture and all training hyper-parameters remained the same. All networks were trained from scratch with randomly initialized weights.

In the first stage of our training scheme, an auto-encoder was trained using an Adam optimizer with $1\text{e-}2$ as learning rate. As a second step, both UNet and shape code discriminator were optimized alternatively at each batch. We used Adam optimizer with $1\text{e-}4$ as learning rate and a regularization weighting factor of $\lambda = 1\text{e-}2$. All networks were trained on 2D slices for $10$ epochs with a batch size of $32$ and extensive data augmentation including random scaling, rotation and shifting. We implemented deep learning models in Keras using a Nvidia RTX 2080 Ti GPU with 12 GB of RAM.

After inference, the predicted 2D masks were stacked into a 3D volume. For each bone, we used connected component analysis to eliminate some of the model outputs and applied 3D morphological closing to smooth the contours.

The performance of each method was evaluated based on the similarity between 3D predicted and ground truth masks. We computed Dice coefficient, relative absolute volume difference (RAVD), average symmetric surface distance (ASSD) and maximum symmetric surface distance (MSSD) metrics for each bone and reported the average scores. All the metrics were determined in a leave-one-out manner.

\begin{table}[t]
\footnotesize
\centering
    \begin{tabular}{|P{.02cm}||P{1.6cm}||P{1.05cm}|P{1.05cm}|P{1.05cm}|P{1.05cm}|} 
     \hline
     & Method & Dice $\uparrow$ & RAVD $\downarrow$ & ASSD $\downarrow$ & MSSD $\downarrow$ \\ 
     \hline\hline
     
     \multirow{4}{*}{\hspace{-0.1cm}\rotatebox[origin=c]{90}{Ankle}} & Base. UNet & 91.9$\pm$3.1 & 9.8$\pm$6.6 & 0.9$\pm$0.7 & 9.1$\pm$6.3 \\\cline{2-6}
     & Adv. Reg. & 91.9$\pm$2.3 & 9.5$\pm$4.8 & 1.0$\pm$0.6 & 9.6$\pm$5.9 \\\cline{2-6}
     & Sh. Reg. & 92.4$\pm$1.9 & 8.9$\pm$5.2 & 0.8$\pm$0.3 & 8.8$\pm$4.4 \\ \cline{2-6}
     & SPAR & \textbf{92.7$\pm$1.3} & \textbf{8.0$\pm$3.6} & \textbf{0.8$\pm$0.2} & \textbf{8.1$\pm$2.9} \\
     \hline\hline
     
     \multirow{4}{*}{\hspace{-0.1cm}\rotatebox[origin=c]{90}{Shoulder}} & Base. UNet & 82.1$\pm$14.4 & 15.4$\pm$22.7 & 2.7$\pm$4.0 & \textbf{23.7$\pm$20.8} \\ \cline{2-6}
     & Adv. Reg. & 82.8$\pm$12.3 & 14.9$\pm$17.7 & 3.4$\pm$4.0 & 29.3$\pm$22.6 \\\cline{2-6}
     & Sh. Reg. & \textbf{83.8$\pm$9.5} & 13.7$\pm$14.7 & \textbf{2.4$\pm$2.5} & 25.8$\pm$20.0 \\ \cline{2-6}
     & SPAR & 83.5$\pm$10.3 & \textbf{13.6$\pm$15.8} & 2.7$\pm$2.7 & 29.1$\pm$22.1 \\
    
     \hline
    \end{tabular}
\caption{Quantitative assessment of UNet regularization methods: baseline UNet \cite{ronneberger_u-net_2015}, adversarial regularization \cite{singh_breast_2020}, shape priors based regularization \cite{oktay_anatomically_2018} and the proposed shape priors based adversarial regularization on ankle and shoulder datasets. Metrics include Dice, RAVD ($\%$), ASSD and MSSD (mm). Best results are in bold.}
\label{table:results}
\end{table}

\subsection{Results}
\label{sec:results}

From the quantitative results (Tab.\ref{table:results}), our method achieved competitive results compared to state-of-the-art on both datasets. On ankle dataset, our approach ranked best in Dice (92.7$\%$), RAVD (8.0$\%$), ASSD (0.8mm) and MSSD (8.1mm) metrics. For shoulder dataset, our method outperformed other approaches in RAVD (13.6$\%$) while remaining second best in Dice (0.3$\%$ lower than the best) and ASSD (0.3mm higher than the best). We suspected that the high variability observed in shoulder results was due to the poor quality of two outlier examinations. The visual comparisons (Fig.\ref{fig:segmentation}) provided the evidence of gradual improvements in segmentation quality of the regularized methods over baseline UNet. We wanted to report statistical significance tests to compare the performance of the employed methods but the required sample size determined using a power analysis (with typical statistical power $\beta = 0.8$) was larger than our available datasets. 

\begin{figure*}[t]
\begin{tikzpicture}

\begin{axis}[name=plot11,
        height=3.5cm,
        width=3.5cm, 
        scatter/classes={
        0={black!40!green},
        1={white!10!blue},
        2={black!15!yellow},
        3={black!20!green},
        4={white!50!blue},
        5={white!10!yellow}},
        grid,
        xmin=-90, 
        xmax=90, 
        ymin=-90,
        ymax=90,
        xtick={-80, -40, 0, 40, 80},
        ytick={-80, -40, 0, 40, 80},
        xticklabels={-80, -40, 0, 40, 80},
        yticklabels={-80, -40, 0, 40, 80},
        ticklabel style = {font=\tiny},
        axis background/.style={fill=gray!10}]
\addplot+[
    scatter src=explicit symbolic,
    only marks,
    mark=*,
    scatter,
    mark size=.8pt
    ]
table[meta=labels]{CAE_a_tSNE_1.dat};
\end{axis}

\begin{axis}[name=plot12,
        at={($(plot11.east)+(.5cm,0)$)},
        anchor=west,
        height=3.5cm,
        width=3.5cm, 
        scatter/classes={
        0={black!40!green},
        1={white!10!blue},
        2={black!15!yellow},
        3={black!20!green},
        4={white!50!blue},
        5={white!10!yellow}},
        grid,
        xmin=-90, 
        xmax=90, 
        ymin=-90,
        ymax=90,
        xtick={-80, -40, 0, 40, 80},
        ytick={-80, -40, 0, 40, 80},
        xticklabels={-80, -40, 0, 40, 80},
        yticklabels={-80, -40, 0, 40, 80},
        ticklabel style = {font=\tiny},
        axis background/.style={fill=gray!10},]
\addplot+[
    scatter src=explicit symbolic,
    only marks,
    mark=*,
    scatter,
    mark size=.8pt
    ]
table[meta=labels]{CAE_a_tSNE_2.dat};
\end{axis}

\begin{axis}[name=plot13,
        at={($(plot12.east)+(.5cm,0)$)},
        anchor=west,
        height=3.5cm,
        width=3.5cm,
        scatter/classes={
        0={black!40!green},
        1={white!10!blue},
        2={black!15!yellow},
        3={black!20!green},
        4={white!50!blue},
        5={white!10!yellow}},
        grid,
        xmin=-90, 
        xmax=90, 
        ymin=-90,
        ymax=90,
        xtick={-80, -40, 0, 40, 80},
        ytick={-80, -40, 0, 40, 80},
        xticklabels={-80, -40, 0, 40, 80},
        yticklabels={-80, -40, 0, 40, 80},
        ticklabel style = {font=\tiny},
        axis background/.style={fill=gray!10}]
\addplot+[
    scatter src=explicit symbolic,
    only marks,
    mark=*,
    scatter,
    mark size=.8pt
    ]
table[meta=labels]{CAE_a_tSNE_3.dat};
\end{axis}

\begin{axis}[name=plot14,
        at={($(plot13.east)+(.5cm,0)$)},
        anchor=west,
        height=3.5cm,
        width=3.5cm, 
        scatter/classes={
        0={black!40!green},
        1={white!10!blue},
        2={black!15!yellow},
        3={black!20!green},
        4={white!50!blue},
        5={white!10!yellow}},
        grid,
        xmin=-90, 
        xmax=90, 
        ymin=-90,
        ymax=90,
        xtick={-80, -40, 0, 40, 80},
        ytick={-80, -40, 0, 40, 80},
        xticklabels={-80, -40, 0, 40, 80},
        yticklabels={-80, -40, 0, 40, 80},
        ticklabel style = {font=\tiny},
        axis background/.style={fill=gray!10}, ]
\addplot+[
    only marks,
    scatter src=explicit symbolic,
    mark=*,
    scatter,
    mark size=.8pt
    ]
table[meta=labels]{CAE_a_tSNE_4.dat};
\end{axis}

\begin{axis}[name=plot15,
        at={($(plot14.east)+(.5cm,0)$)},
        anchor=west,
        height=3.5cm,
        width=3.5cm, 
        scatter/classes={
        0={black!40!green},
        1={white!10!blue},
        2={black!15!yellow},
        3={black!20!green},
        4={white!50!blue},
        5={white!10!yellow}},
        grid,
        xmin=-90, 
        xmax=90, 
        ymin=-90,
        ymax=90,
      xtick={-80, -40, 0, 40, 80},
        ytick={-80, -40, 0, 40, 80},
        xticklabels={-80, -40, 0, 40, 80},
        yticklabels={-80, -40, 0, 40, 80},
        ticklabel style = {font=\tiny},
        axis background/.style={fill=gray!10}, ]
\addplot+[
    only marks,
    scatter src=explicit symbolic,
    mark=*,
    scatter,
    mark size=.8pt
    ]
table[meta=labels]{CAE_a_tSNE_5.dat};
\end{axis}

\begin{axis}[name=plot16,
        at={($(plot15.east)+(.5cm,0)$)},
        anchor=west,
        height=3.5cm,
        width=3.5cm, 
        scatter/classes={
        0={black!40!green},
        1={white!10!blue},
        2={black!15!yellow},
        3={black!20!green},
        4={white!50!blue},
        5={white!10!yellow}},
        grid,
        xmin=-90, 
        xmax=90, 
        ymin=-90,
        ymax=90,
      xtick={-80, -40, 0, 40, 80},
        ytick={-80, -40, 0, 40, 80},
        xticklabels={-80, -40, 0, 40, 80},
        yticklabels={-80, -40, 0, 40, 80},
        ticklabel style = {font=\tiny},
        axis background/.style={fill=gray!10}, ]
\addplot+[
    only marks,
    scatter src=explicit symbolic,
    mark=*,
    scatter,
    mark size=.8pt
    ]
table[meta=labels]{CAE_a_tSNE_10.dat};
\end{axis}

\begin{axis}[name=plot21,
        at={($(plot11.south)-(0,.5cm)$)},
        anchor=north,
        height=3.5cm,
        width=3.5cm, 
        scatter/classes={
        0={black!40!green},
        1={white!10!blue},
        2={black!15!yellow},
        3={black!20!green},
        4={white!50!blue},
        5={white!10!yellow}},
        grid,
        xmin=-90, 
        xmax=90, 
        ymin=-90,
        ymax=90,
        xtick={-80, -40, 0, 40, 80},
        ytick={-80, -40, 0, 40, 80},
        xticklabels={-80, -40, 0, 40, 80},
        yticklabels={-80, -40, 0, 40, 80},
        ticklabel style = {font=\tiny},
        axis background/.style={fill=gray!10}, ]
\addplot+[
    only marks,
    scatter src=explicit symbolic,
    mark=*,
    scatter,
    mark size=.8pt
    ]
table[meta=labels]{LD_a_tSNE_1.dat};
\end{axis}

\begin{axis}[name=plot22,
        at={($(plot21.east)+(.5cm,0)$)},
        anchor=west,
        height=3.5cm,
        width=3.5cm, 
        scatter/classes={
        0={black!40!green},
        1={white!10!blue},
        2={black!15!yellow},
        3={black!20!green},
        4={white!50!blue},
        5={white!10!yellow}},
        grid,
        xmin=-90, 
        xmax=90, 
        ymin=-90,
        ymax=90,
        xtick={-80, -40, 0, 40, 80},
        ytick={-80, -40, 0, 40, 80},
        xticklabels={-80, -40, 0, 40, 80},
        yticklabels={-80, -40, 0, 40, 80},
        ticklabel style = {font=\tiny},
        axis background/.style={fill=gray!10}, ]
\addplot+[
    only marks,
    scatter src=explicit symbolic,
    mark=*,
    scatter,
    mark size=.8pt
    ]
table[meta=labels]{LD_a_tSNE_2.dat};
\end{axis}

\begin{axis}[name=plot23,
        at={($(plot22.east)+(.5cm,0)$)},
        anchor=west,
        height=3.5cm,
        width=3.5cm, 
        scatter/classes={
        0={black!40!green},
        1={white!10!blue},
        2={black!15!yellow},
        3={black!20!green},
        4={white!50!blue},
        5={white!10!yellow}},
        grid,
        xmin=-90, 
        xmax=90, 
        ymin=-90,
        ymax=90,
        xtick={-80, -40, 0, 40, 80},
        ytick={-80, -40, 0, 40, 80},
        xticklabels={-80, -40, 0, 40, 80},
        yticklabels={-80, -40, 0, 40, 80},
        ticklabel style = {font=\tiny},
        axis background/.style={fill=gray!10}, ]
\addplot+[
    only marks,
    scatter src=explicit symbolic,
    mark=*,
    scatter,
    mark size=.8pt
    ]
table[meta=labels]{LD_a_tSNE_3.dat};
\end{axis}

\begin{axis}[name=plot24,
        at={($(plot23.east)+(.5cm,0)$)},
        anchor=west,
        height=3.5cm,
        width=3.5cm, 
        scatter/classes={
        0={black!40!green},
        1={white!10!blue},
        2={black!15!yellow},
        3={black!20!green},
        4={white!50!blue},
        5={white!10!yellow}},
        grid,
        xmin=-90, 
        xmax=90, 
        ymin=-90,
        ymax=90,
        xtick={-80, -40, 0, 40, 80},
        ytick={-80, -40, 0, 40, 80},
        xticklabels={-80, -40, 0, 40, 80},
        yticklabels={-80, -40, 0, 40, 80},
        ticklabel style = {font=\tiny},
        axis background/.style={fill=gray!10}, ]
\addplot+[
    only marks,
    scatter src=explicit symbolic,
    mark=*,
    scatter,
    mark size=.8pt
    ]
table[meta=labels]{LD_a_tSNE_4.dat};
\end{axis}

\begin{axis}[name=plot25,
        at={($(plot24.east)+(.5cm,0)$)},
        anchor=west,
        height=3.5cm,
        width=3.5cm, 
        scatter/classes={
        0={black!40!green},
        1={white!10!blue},
        2={black!15!yellow},
        3={black!20!green},
        4={white!50!blue},
        5={white!10!yellow}},
        grid,
        xmin=-90, 
        xmax=90, 
        ymin=-90,
        ymax=90,
        xtick={-80, -40, 0, 40, 80},
        ytick={-80, -40, 0, 40, 80},
        xticklabels={-80, -40, 0, 40, 80},
        yticklabels={-80, -40, 0, 40, 80},
        ticklabel style = {font=\tiny},
        axis background/.style={fill=gray!10}, ]
\addplot+[
    only marks,
    scatter src=explicit symbolic,
    mark=*,
    scatter,
    mark size=.8pt
    ]
table[meta=labels]{LD_a_tSNE_5.dat};
\end{axis}

\begin{axis}[name=plot26,
        at={($(plot25.east)+(.5cm,0)$)},
        anchor=west,
        height=3.5cm,
        width=3.5cm, 
        scatter/classes={
        0={black!40!green},
        1={white!10!blue},
        2={black!15!yellow},
        3={black!20!green},
        4={white!50!blue},
        5={white!10!yellow}},
        grid,
        xmin=-90, 
        xmax=90, 
        ymin=-90,
        ymax=90,
        xtick={-80, -40, 0, 40, 80},
        ytick={-80, -40, 0, 40, 80},
        xticklabels={-80, -40, 0, 40, 80},
        yticklabels={-80, -40, 0, 40, 80},
        ticklabel style = {font=\tiny},
        axis background/.style={fill=gray!10}, ]
\addplot+[
    only marks,
    scatter src=explicit symbolic,
    mark=*,
    scatter,
    mark size=.8pt
    ]
table[meta=labels]{LD_a_tSNE_10.dat};
\end{axis}

\node[anchor=south] at ($(plot11.north)$) {\scalebox{.9}{Epoch 1}};
\node[anchor=south] at ($(plot12.north)$) {\scalebox{.9}{Epoch 2}};
\node[anchor=south] at ($(plot13.north)$) {\scalebox{.9}{Epoch 3}};
\node[anchor=south] at ($(plot14.north)$) {\scalebox{.9}{Epoch 4}};
\node[anchor=south] at ($(plot15.north)$) {\scalebox{.9}{Epoch 5}};
\node[anchor=south] at ($(plot16.north)$) {\scalebox{.9}{Epoch 10}};

\node[rotate=90] at ($(plot11.west)+(-.6cm,0)$) {\scalebox{.9}{Sh. Reg.}};
\node[rotate=90] at ($(plot21.west)+(-.6cm,0)$) {\scalebox{.9}{SPAR}};

\draw[line width=0.01mm] (1.7,-3) -- (1.7,-3.5) -- (12.3,-3.5) -- (12.3,-3) -- cycle;
\node[anchor=west] at (1.7,-3.25) {\scalebox{.9}{\underline{Ground truth:}}};
\node[anchor=west] at (7.2,-3.25) {\scalebox{.9}{\underline{Prediction:}}};

\node[anchor=west] at (3.7,-3.25) {\scalebox{.9}{Calcaneus}};
\node[anchor=west] at (5.25,-3.25) {\scalebox{.9}{Talus}};
\node[anchor=west] at (6.2,-3.25) {\scalebox{.9}{Tibia}};

\fill[black!40!green] (3.7,-3.25) circle (.075);
\fill[white!10!blue] (5.25,-3.25) circle (.075);
\fill[black!15!yellow] (6.2,-3.25) circle (.075);

\node[anchor=west] at (8.85,-3.25) {\scalebox{.9}{Calcaneus}};
\node[anchor=west] at (10.4,-3.25) {\scalebox{.9}{Talus}};
\node[anchor=west] at (11.35,-3.25) {\scalebox{.9}{Tibia}};

\fill[black!20!green] (8.85,-3.25) circle (.075);
\fill[white!50!blue] (10.4,-3.25) circle (.075);
\fill[white!10!yellow] (11.35,-3.25) circle (.075);

\end{tikzpicture}
\caption{Visualization of the convergence in latent space of the shape priors based regularization and the proposed shape priors based adversarial regularization methods on the ankle dataset. This visualization was obtained using the t-SNE algorithm in which each colored dot corresponds to a 2D binary mask (ground truth or prediction) of one of the three structures of interest (calcaneus, talus and tibia).}
\label{fig:t_sne_viz}
\end{figure*}
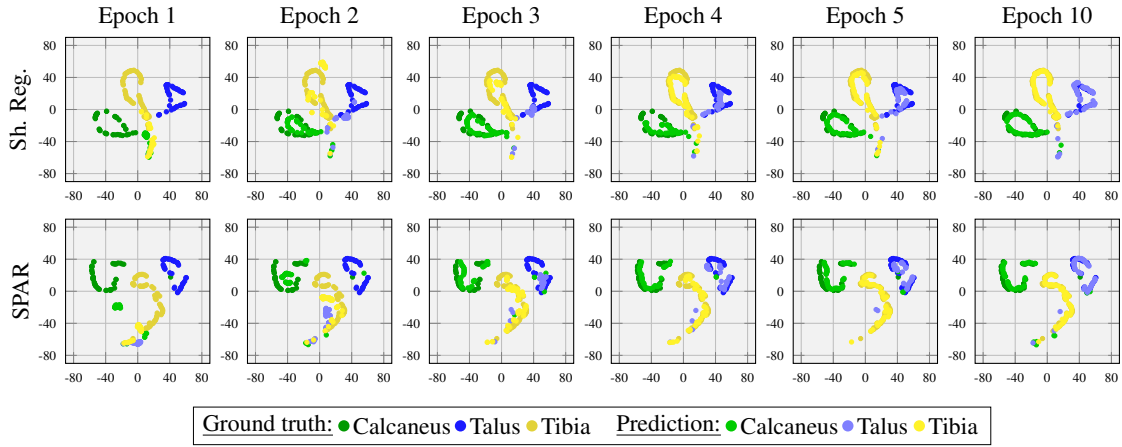{}

\subsection{Convergence in latent shape space}
\label{sec:latent_space_visualization}

We compared the convergence in latent shape space of the shape priors based regularization \cite{oktay_anatomically_2018} and our proposed shape priors based adversarial regularization. The latent representation of the predicted segmentation masks was visualized from epoch 1 to 5 and 10. We used the shape encoder to project the predicted and ground truth masks into the latent shape space and applied global max-pooling to obtain 512 dimensional codes for each 2D bone mask. Finally, in order to visualize the collected codes, we employed the t-SNE algorithm to embed the data into a 2D space (Fig.\ref{fig:t_sne_viz}).

Visualizing the latent shape space provided a qualitative validation of the added shape priors as predicted segmentations converged towards ground truths. However, it would also be useful to analyse outliers and their corresponding segmentation masks to further improve training.

\section{Conclusion}
\label{sec:conclusion}

This study presented a deep-learning based regularization scheme with promising results for the task of multi-structure bone segmentation in MR images. Based on a non-linear representation of the shape, the proposed strategy exploited a shape code discriminator to enforce globally consistent shape predictions. Future work is aimed at integrating muscles into the segmentation framework, which could help better analyse bone-muscle interactions during child development.

\small{
\section{Compliance with Ethical Standards}
\label{sec:ethics}

MRI data acquisition on the pediatric cohort used in this study was performed in line with the principles of the Declaration of Helsinki. Ethical approval was granted by the Ethics Committee (Comit\'e Protection de Personnes Ouest VI) of CHRU Brest (2015-A01409-40).

\section{Acknowledgments}
\label{sec:acknowledgments}

This work was funded by IMT, Fondation Mines-Télécom and Institut Carnot TSN (Futur \& Ruptures). Data were acquired with the support of Fondation motrice (2015/7), Fondation de l’Avenir (AP-RM-16-041), PHRC 2015 (POPB 282) and Innoveo (CHRU Brest).

\bibliographystyle{IEEEbib}
\bibliography{ISBI2021}}

\end{document}